# Discovery of switchable weak topological insulator state in quasi-one-dimensional bismuth iodide

R. Noguchi,[1] T. Takahashi,[2] K. Kuroda,[1] M. Ochi,[3] T. Shirasawa,[4] M. Sakano,[1,5]
C. Bareille,[1] M. Nakayama,[1] M. D. Watson,[6] K. Yaji,[1] A. Harasawa,[1] H. Iwasawa,[6]
P. Dudin,[6] T. K. Kim,[6] M. Hoesch,[6] S. Shin,[1] R. Arita,[7] T. Sasagawa,[2] and Takeshi Kondo*[1]

[1]*Institute for Solid State Physics, University of Tokyo, Kashiwa, Chiba 277-8581, Japan*

[2]*Materials and Structures Laboratory, Tokyo Institute of Technology,*
*Yokohama, Kanagawa 226-8503, Japan*

[3]*Department of Physics, Osaka University, Toyonaka, Osaka 560-0043, Japan*

[4]*National Metrology Institute of Japan,*
*National Institute of Advanced Industrial Science*
*and Technology, Tsukuba, Ibaraki 305-8565, Japan*

[5]*Department of Applied Physics and Quantum-Phase Electronics Center (QPEC),*
*The University of Tokyo, Tokyo 113-8656, Japan*

[6]*Diamond Light Source, Harwell Campus,*
*Didcot, OX11 0DE, United Kingdom*

[7]*RIKEN Center for Emergent Matter Science (CEMS), Wako, Saitama 351-0198, Japan*

(Dated: February 9, 2018)

The major breakthroughs in the understanding of topological materials over the past decade were all triggered by the discovery of the $Z_2$ topological insulator (TI). In three dimensions (3D), the TI is classified as either "strong" or "weak" [1, 2], and experimental confirmations of the strong topological insulator (STI) rapidly followed the theoretical predictions [3–5]. In contrast, the weak topological insulator has so far eluded experimental verification, since the topological surface states emerge only on particular side surfaces which are typically undetectable in real 3D crystals [6–10]. Here we provide experimental evidence for the WTI state in a bismuth iodide, $\beta$-Bi4I4. Significantly, the crystal has naturally cleavable top and side planes both stacked via van-der-Waals forces, which have long been desirable for the experimental realization of the WTI state [11, 12]. As a definitive signature of it, we find quasi-1D Dirac TSS at the side-surface (100) while the top-surface (001) is topologically dark. Furthermore, a crystal transition from the $\beta$- to $\alpha$-phase drives a topological phase transition from a nontrivial WTI to the trivial insulator around room temperature. This topological phase, viewed as quantum spin Hall (QSH) insulators stacked three-dimensionally [13, 14], and excellent functionality with on/off switching will lay a foundation for new technology benefiting from highly directional spin-currents with large density protected against backscattering.

The quasi-1D compounds $\alpha$-Bi$_4$I$_4$ and $\beta$-Bi$_4$I$_4$ share similar crystal structures, formed from arrangements of Bi$_4$I$_4$ chains within the space group C2/$m$ (No. 12) [15]. They differ only in their stacking sequences along the $c$-axis as shown in Figs. 1a and 1b. The unit cell of the $\beta$-phase consists of a single Bi$_4$I$_4$ block, while the $\alpha$-phase has a different stacking of double Bi$_4$I$_4$ blocks along the $c$ axis, leading to a larger cell. Despite the small difference between these two crystal structures, distinct transport properties are obtained: the $\alpha$-phase exhibits a typical semiconductor-like resistivity, whereas the $\beta$-phase, in contrast, presents conductive behavior (Fig. 1c). A crystal phase transition with a hysteresis is observed in the resistivity around room temperature while the temperature is slowly swept at a rate of 3 K/min (Fig. 1d). However, since the high temperature phase (the $\beta$-phase) can be pinned by quenching crystals (Method, Supplementary information), both phases can be equally investigated at low temperatures.

The band topology in the $\beta$-phase is calculated in Figs. 1f and 1g. The bulk band struc-ture without spin-orbit coupling (SOC) has small gaps ($E_G$) at the L and M points between the conduction and valence bands with opposite parities derived from Bi 6$p$ orbitals (Fig. 1f). Inclusion of SOC induces parity inversions leading to different topological phases. The $Z_2$ topological invariants ($\nu_0$:$\nu_1\nu_2\nu_3$) are obtained with the method of Fu and Kane [2], accounting for the band inversions at time-reversal invariant momenta (TRIMs) in 3D Brillouin zone (BZ) (Fig. 1e). We note, however, that the $Z_2$ invariants cannot be conclusively determined only by the DFT calculations, since $\beta$-Bi$_4$I$_4$ is in proximity of three different phases [11, 12]: a STI (1;110), a WTI (0;001), or even a trivial phase (0;000), sensitively depending on correction of $E_G$ (Fig. 1g). Experimental determination is therefore required through the direct observation of relevant TSS. In contrast to the uncertainty in $\beta$-Bi$_4$I$_4$, $\alpha$-Bi$_4$I$_4$ is constrained to being trivial regardless of gap size, since its crystal structure with double blocks ensures even-numbered parity inversions at every TRIM (Supplementary information); however this too should be confirmed by experiments.

The STI state manifests with TSS at all of the surfaces (Figs. 1j and 1k), as has been shown in a number of Bi-based chalcogenides [16–18]. In contrast, TSS should emerge only on particular surfaces in the WTI state [1, 2], which are topologically protected [19–21]. The DFT calculations for the WTI state in $\beta$-Bi$_4$I$_4$ predict that the top-surface (001) is topologically dark with an absence of TSS (Supplementary information), and instead the quasi-1D Dirac topological SS emerges at the side-surface (100) (Fig. 1i) [22]. This material state is thus regarded as the 3D stacking of 2D QSH states as schematically shown in Fig. 1h. Notably, the quasi-1D Dirac-state exhibits a characteristic spin texture in $k$-space, which differs from the helical spin texture of the 2D Dirac TSS in the STIs (Fig. 1k): the spin direction is locked almost parallel along the quasi-1D Fermi surface (FS) and antiparallel between $\pm k_z$ (Supplementary information). This spin-texture most likely prohibits the electron-backscattering, thus generating robust spin-current, similar to the edge state in a 2D-QSH insulator [13, 14].

In contrast to the 2D Dirac TSS of the STI state with a large gap opened at the $\bar{Z}$ point (Figs. 1n and 1o), the quasi-1D Dirac TSS for the side-plane (100) of WTI features gapless states not only at $\bar{\Gamma}$ but also at $\bar{Z}$ (Figs. 1l and 1m). This clear difference could definitively distinguish the WTI state from the STI state by experiments. Therefore we conclude that systematic investigations of not only the top-surface (001) [12] but also the side-surface (100) are the only way to unambiguously identify the topological nature of this material.

Using surface-sensitive techniques, we demonstrate that the quasi-1D crystal naturally exposes the (001) and (100) planes by cleaving the *ab*-plane, which is largest in the grown crystals (Fig. 2a). The top-surface (001) is seen as terraces in images taken by scanning electron microscope and laser-microscope (Figs. 2b and 2c, respectively). A number of facets oriented along the 1D-chain direction can instead be assigned as the side-surface (100). This has been indeed confirmed by grazing-incidence small angle X-ray scattering (GISAXS) measurement of the cleaved surface as shown in Fig. 2d; we find two strong X-ray reflections coming from the terraces and facets. The angle between the two streaks is estimated to be $\sim 72°$, which is consistent with the angle between the (001) and (100) planes of $\beta$-$Bi_4I_4$ (Fig. 1b) determined by powder X-ray diffraction measurements [15]. $Bi_4I_4$ with an experimentally accessible side-surface should be discriminated from the other candidate materials for WTI, which do not have such a surface though essential to determine the novel topological phase. This compound therefore provides a long desired platform to experimentally realize and investigate a WTI.

To determine the electronic structures of both $\alpha$-$Bi_4I_4$ and $\beta$-$Bi_4I_4$, we have performed laser-ARPES measurement with high energy and momentum resolutions. The two crystallographic phases were selectively obtained by the quench method (Methods). The ARPES signals from the (001) and (100) planes are accumulated all together in the experiments, since the incident laser light with $\sim 50$ $\mu$m in spot size illuminates both the terraces and facets exposed on the cleaved surface (Figs. 2b,c and 3a). The mapping of photoelectron intensities at the Fermi energy ($E_F$) exhibits clearly different FS structures for the two materials (Figs. 3b and 3c): a quasi-1D like shape is obtained in $\beta$-phase, whereas the island-like intensities are detected only around $\bar{M}$ in BZ of the (001) plane for the $\alpha$-phase.

In $\alpha$-$Bi_4I_4$, we observe trivial band structures as expected by the theoretical calculations. Figures 3d-g display the ARPES band maps in the $\alpha$-phase at different emission angles ($\theta$) with respect to the normal of the top-surface (001) (Fig. 3a). The valence band shows a strong energy dispersion along $k_x$, which varies with $\theta$-rotation (Figs. 3d-g). The top of the valence band is found at $\theta=36°$, which corresponds to the $\bar{M}$ point in BZ for the (001) plane (Fig. 3b). The conduction band is slightly occupied in the data mostly due to the surface band bending, which generates island-like intensities in FS mapping (Fig. 3b). No in-gap Dirac-state is observed at the $\bar{\Gamma}$ (Fig. 3d) and $\bar{M}$ points (Fig. 3g), indicating trivial band topology.

In contrast to the case of $\alpha$-phase, we find a Dirac-cone like energy dispersion near $E_\text{F}$ in the $\beta$-phase (Figs. 3h-m). This band has a quasi 1D character as already seen in Fig. 3c, and it is only weakly dispersive in energy along $k_x$ (or with $\theta$-variation). Since this anomalous state is not detected in the trivial $\alpha$-Bi$_4$I$_4$, it should be due to a topological surface. A similar quasi-1D state was also confirmed by ARPES at a higher photon energy of 85 eV (Supplementary information), although the intensities are extremely weak and thus barely detectable [12]. Surprisingly, the Dirac-state is observed even at $\theta=0°$, or $\bar{\Gamma}$ in the BZ of the (001) surface, whereas no such state is predicted by the DFT calculations for any cases of band topology (Supplementary information). The only possible explanation for it is that the observed quasi-1D Dirac-state is derived from the side-surface (100).

The nontrivial band topology of either STI or WTI can be determined by comparing the observed surface dispersions with theoretical expectations in Figs. 1l-o: a large gap should be observed for a STI at the $\bar{\text{Z}}$ point, while no gap is expected for a WTI. In Figs. 3l and 3m, we show the ARPES band maps obtained at $\theta=49°$ and 72°, which cut across the $\bar{\text{Z}}$ and $\bar{\Gamma}$ points along $k_y$, respectively, in the BZ for the (100) surface (see red and yellow lines in Fig 3c). The observed Dirac dispersions are gapless at both the high-symmetry $k$ points. This is fully consistent with the DFT calculation for the WTI state (Figs. 1l and 1m), thus our ARPES results validate the classification of $\beta$-Bi$_4$I$_4$ as a WTI.

Our arguments is however still left with an ambiguity that needs to be resolved before finalizing the conclusion. So far, the band structures for the key (100) plane have been contaminated by signatures from the (001) plane in the data measured by laser-ARPES. The spot size of incident laser-beam ($\sim$50 $\mu$m) was much larger than the step height of the facets (typically $\sim$2 $\mu$m; see Fig 2), which has prevented us from detecting the pure signatures only from the (100) plane. To address this issue and exclusively examine the WTI surface, we have utilized a surface-selective ARPES, so-called nano-ARPES, equipped with a zone plate to focus the photon beam to a spot less than 1 $\mu$m in size, and selectively observed only the (100) plane without any contamination.

Figure 4a show the experimental geometry for the surface selective investigation on the side-surface (100). Here we set up the thin crystal and cleaved the side-plane ($bc$-plane) with a very narrow width of $<$ 50 $\mu$m. As shown in Figs. 4b, the microscopic intensity map of nano-ARPES captures the property of the the side-surface (100). We shone a light on a part of bright region (white circle in Fig. 4b), which displays relatively uniform intensities indicative of a very flat surface within ∼10×100 $\mu$m. This surface-selective experiment confirms the quasi-1D FS (Fig 4d) expected for a WTI. The relevant linear dispersions due to the TSS is clearly displayed in Figs. 4g and 4, which plots the band maps measured by nano-ARPES along the high-symmetry momentum lines across $\bar{\Gamma}$ and $\bar{Z}$ (red and yellow lines in Figs. 4d). For comparison, we have also measured the top-surface (001) using nano-ARPES, with the experimental geometry presented in 4c. In contrast to the side-surface (100), no quasi-1D TSS is observed in the topologically dark surface (Figs. 4j and Figs. 4k). All of these results therefore show a good agreement with our conclusion about the WTI state in $\beta$-$Bi_4I_4$ reached through the measurements with a laser-ARPES.

The WTI state we have revealed may provide various scientific and technological prospects. Since this fascinating state is regarded as the 3D analogue of the QSH insulator, which could generate highly directional spin-current over a wide surface, our discovery will stimulate further in-depth study of exotic quantum phenomena [22–25]. In particular, $Bi_4I_4$ has an excellent functionality that the emergence of robust spin currents [13, 14] can be controlled around room temperature by selecting the crystal phases which are either topological or non-topological. Our experimental observations thus initiate the basic and technological researches on the 3D analogue of a QSH insulator, which may well lead to novel electronic and spintronic technologies for future applications.

---

**Methods Samples.**

Single crystals of $Bi_4I_4$ were grown by the chemical vapour transport (CVT) method, similar to the Ref. [26]. Bi and $HgI_2$ at a molar ratio of 1 : 1 were sealed into a quartz tube under vacuum. The quartz tube was placed into a home-built transparent two-heating-zone furnace, which was described in Ref. [27]. By real-time observation, the growth temperatures were optimized to be 285°C at the source zone and 188°C at the nucleation zone, respectively. After 72 hours of the growth, crystals in the shape of rectangular sheets with typical dimensions ∼2.0×0.5×0.03 mm$^3$ were successfully obtained. The single-crystal X-ray diffraction measurement indicated the middle, the longest, and the shortest edges were along $a$, $b$, and $c$ axes, respectively.

We found that the $\alpha$-phase ($\beta$-phase) was thermodynamically stable below (above) 300 K. These two phases were distinguishable by the X-ray diffraction patterns from the $ab$-plane of the single crystals (Supplementary information); only the $\alpha$-phase showed a peak (i.e. 005-diffraction) at 22 deg. By a rapid quenching from 400 K to the liquid nitrogen temperature (77 K), the structural transition from the $\beta$-phase to the $\alpha$-phase was found to be suppressed, thereby it was possible to characterize the $\beta$-phase at low temperatures by resistivity and ARPES measurements.

**Transport measurements.**
Applying the appropriate temperature cycle, the electrical resistivity along the $b$-axis of both the $\alpha$- and $\beta$-phases was measured by a standard four probe technique in the same crystal with the same electrical contacts. The results obtained by this procedure are shown in Figs. 1c and 1d, demonstrating their different temperature dependence of the resistivity and the crystal phase transition around room temperature.

**DFT calculations.**
Within the DFT framework, first-principles electronic-states calculations were performed using the full-potential linearized augmented plane-wave method as implemented in the WIEN2k code [28]. In-addition to the generalized gradient approximation of Perdew, Burke and Ernzerhof (GGA-PBE) of exchange-correlation functional [29], the modified Becke-Johnson exchange potential (mBJ) [30] with manual adjustment of two parameters was used to see the effect of the gap size correction as presented in Figs. 1f and 1g. Spin-orbit coupling (SOC) was included as a second variational step with a basis of scalar-relativistic eigenfunctions. The experimentally determined lattice constants were used for all the calculations. The plane-wave cutoff energy was set to $R_{MT}K_{max} = 7$, where the muffin tin radii were $R_{MT} = 2.5$ a.u. for both Bi and I. The Brillouin zone was sampled with the Monkhorst-Pack scheme [31] with momentum grids finer than $\Delta k = 0.012$ Å$^{-1}$ (i.e. 20 x 8 x 20 k-point mesh for $\beta$-Bi$_4$I$_4$). From the calculated band structures, we extracted the Wannier functions of the Bi- and I-p orbitals using the Wien2wannier [32] and Wannier90 [33] codes. The semi-infinite-slab tight-binding models constructed from these Wannier functions were used for calculating the surface spectra in the way described in Ref. [34].

**GISAXS experiments**

GISAXS experiments were performed at beamline 18B of the Photon Factory at KEK. The X-ray energy was 10.8 keV, and the slit-cut beam size was 0.1×0.1 mm. The glancing angle of the X-rays was 0.1° with respect to the $\beta$-Bi$_4$I$_4$ (001) plane as shown in Fig. 2c. The scattered X-rays were detected with a PILATUS 1M located 1.2 m downstream of the sample. For the alignment of sample orientation, the 001 and 200 Bragg reflections were used. All measurements were performed at room temperature. The measured surface of the crystal was obtained by cleavage.

**ARPES set-up.**

Laser-based ARPES measurements were performed at the Institute for Solid State Physics, The University of Tokyo [35]. The laser system provides 6.994-eV photons [36]. Photoelectrons were analysed with a ScientaOmicron DA30L analyzer. The measurement sample temperature was about 20 K. The angle resolution was 0.3° and the overall energy resolution was set to less than 5 meV. Synchrotron-based nano-ARPES measurements were performed at nano-ARPES branch of the beamline I05 of the Diamond Light Source equipped with a ScientaOmicron DA30L analyzer. The measurement sample temperature was about 30 K. The angle resolution was 0.2° and the overall energy resolutions were better than 60 meV. Note that the new type of photoelectron analyzer is equipped with a electron deflector system, which gives a capability to perform Fermi surface map without the sample rotation. This was greatly advantageous for our experiments as it allows the measurement of quasi-1D band dispersions from a single domain, and thus to separately identify the (001) and (100) signals.

Standard-ARPES measurement with synchrotron radiation was performed at HR-ARPES branch of the beamline I05 of the Diamond Light Source equipped with a Scienta R4000 analyzer. The sample temperature was about 10 K. The angular resolution was 0.2° and the overall energy resolution was better than 20 meV.

**Fermi surface mapping with laser-ARPES**

In Figs. 3b and 3c, we present the Fermi surface mapping on (001) and (100) surfaces scanned by low-energy laser-ARPES. To obtain these results, we collected the photoelectron intensities at different angles ($\theta$) with respect to the normal to the top-surface (001) as shown in Fig. 3a. In this geometry, our ARPES image can cut along the high-symmetry line cross the $\bar{\Gamma}$ point of the (001) surface BZ at $\theta=0°$ and the distinct $\bar{\Gamma}$ point of the (100) surface BZ at $\theta=72°$, respectively. To calculate the relation of the momentum information on the (100) surface BZ and the $\theta$, we use the work function for the (001) surface estimated by the FS mapping as shown in Fig. 3c.

**The bulk band structures and the band topology in $\alpha$-Bi$_4$I$_4$.**

In Fig. 1f and 1g, we theoretically investigate the bulk band dispersions and the band topology with different $E_G$ corrections. We present the theoretical bulk band structures of $\alpha$-phase in Supplementary information. In contrast to the $\beta$-phase, we observe the splitting in both the conduction and valence bands (Supplementary information). This splitting is explained by "bilayer-splittings" due to the conventional unit of the $\alpha$-phase, which consists of the two Bi$_4$I$_4$ blocks along the $c$-axis (see Figs. 1a and 1b). In accordance with the theory, we observe the bilayer splitting in the $\alpha$-phase but it is absent in the $\beta$-phase (Supplementary information). As an important consequence of the bilayer splitting, the number of parity inversions is always restricted to be even at each TRIM. Therefore, the $\alpha$-phase should be a trivial insulator.

**ARPES measurement with synchrotron radiation for an observation of the quasi-1D state in $\beta$-Bi$_4$I$_4$.**

The quasi-1D surface-state at the side-surface (100) of $\beta$-Bi$_4$I$_4$ is clearly observed by high-resolution ARPES (Fig. 3z), despite the top-surface (001) being predominant on the cleaved surface. This comes from the facts that the incident light also illuminates the side-surface (100) which appeared as the facet (Fig. 2). To validate this conclusion, we further performed different ARPES measurement with synchrotron radiation. For this measurement, we choose the photon energy of 85 eV [12]. The additional data are presented in Supplementary information. To correct the photoelectron intensity at different emission angle ($\theta$) with respect to the normal of the top-surface (001), we rotate the sample (Supplementary information).

In the FS maps in Supplementary information, we observe similar FS structures for $\alpha$- and $\beta$-Bi$_4$I$_4$: small electron pockets appear at $\bar{M}$ point of the (001) surface BZ. The overall periodicity seen in the observed dispersions along $k_y$ confirms the cleaved (001) surface.

However, we see the weak but clear signals of the quasi-1D state only in the $\beta$-phase, which should come from the side-surface. This is demonstrated in Supplementary information. Apparently, in contrast to the $\alpha$-phase, the data for $\beta$-phase presents steep dispersion seen in the bulk band gap even at $\theta=0°$ which corresponds to $\bar{\Gamma}$ of the (001) surface. These signals are considered to arise from the quasi-1D surface states of the (100) surface, which shows good agreement with the results for high-resolution laser-ARPES.

---

**Acknowledgements**

We thank Y. Okada, M. Lippmaa, Y. Yoshida, T. Miyamachi, T. Hattori, Y. Hasegawa, and F. Komori for AFM/STM characterization of $Bi_4I_4$ surface and fruitful discussions, and Y. Ishida for supporting analysis of data. This work is supported by Japan Science and Technology Agency, Grants-in-Aid for Scientific Research on Innovative Areas 'J-Physics' (Grant Nos. 15H05852 and 15H05883), and 'Topological Materials Science' (Grant No. 16H00979), and Grants-in-Aid for Young Scientists A (Grants No. 16H06013) and B (Grants No. 17K14319). We thank Diamond Light Source for access to beamline I05 under proposal SI15095, SI16161 and SI17816 that contributed to the results presented here. The GISAXS experiments were performed under the approval of PF-PAC No. 2016G548. The work done at Tokyo Tech. was supported by a JST-CREST project [JPMJCR16F2] and a JSPS Grants-in-Aid for Scientific Research (B) [16H03847]. R. N. was supported by JSPS through Program for Leading Graduate Schools (ALPS).

**Author Contribution**

T.K. and T.S. planned the experimental project. R.N. and K.K. conducted ARPES experiment and analyzed the data. M.S., C.B., M.N., M.D.W., K.Y., A.H., H.I., P.D., T.K.K., M.H., S.S., and T.K. supported ARPES experiment. T.T. and T.S. made and characterized $\alpha$- and $\beta$-$Bi_4I_4$ single crystals, measured laser-microscope image, and performed transport experiments. T.S. performed GISAXS experiment. R.N. measured SEM image. T.S., M.O. and R.A. calculated the band structure and analyzed the band topology. R.N., K.K., and T.K. wrote the paper. All authors discussed the results and commented on the manuscript.

**Additional information**

Competing financial interests: The authors declare no competing financial interests.

**Correspondence**

Correspondence and requests for materials should be addressed to T.K. (email: kondo1215@issp.u-tokyo.ac.jp).

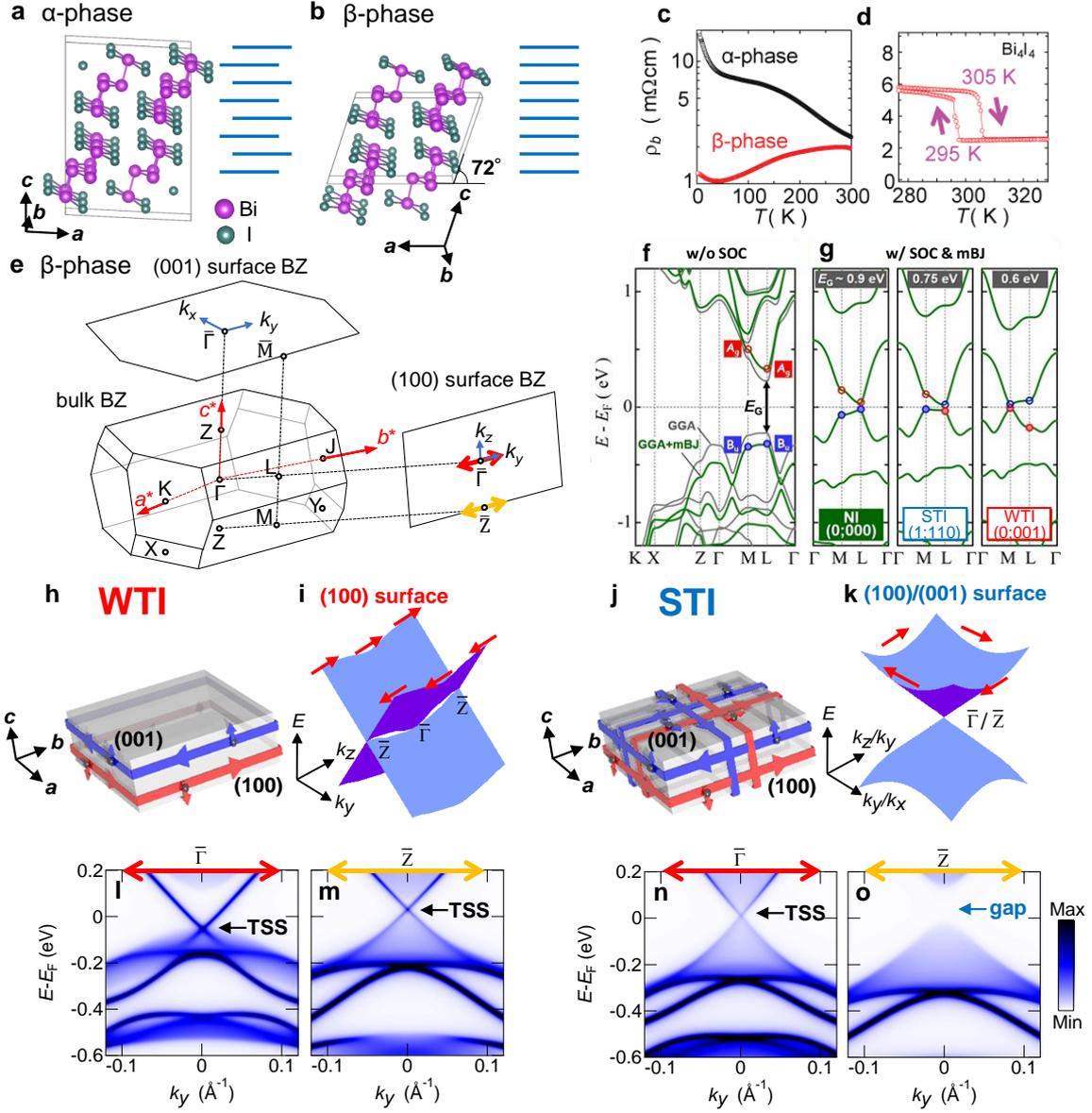

FIG. 1: **Crystal structure, transport and band topology calculations of $\alpha$- and $\beta$-Bi$_4$I$_4$.**
**a, b,** Crystal structures of the $\alpha$-phase (**a**) and the $\beta$-phase (**b**) viewed from the chain direction along $b$ axis. **c,d,** Temperature dependence of their electric resistivity along the chain direction ($\rho_b$). In **d,** the hysteresis was observed by changing the crystal temperature mildly with 0.3 K/min, displaying the phase transition between the $\alpha$- and $\beta$-Bi$_4$I$_4$. **e,** The bulk and projected surface Brillouin zone (BZ) of the $\beta$-phase. **f,** DFT calculations for the $\beta$-phase without the spin-orbit coupling (SOC) obtained by (gray) GGA approximation and (green) GGA+mBJ approximation [Methods]. **g, h,** Band inversions of the $\beta$-phase with inclusion of SOC. The red and blue circles label the even and odd parities, respectively, for the bulk bands at M and L points. The gap size was adjusted by varying parameters in mBJ exchange potential. **h-k,** Schematics of the topological surface states (TSS) of the $\beta$-Bi$_4$I$_4$ in the real space and their band dispersions in reciprocal space for the WTI phase (**h, i**) and the STI phase (**j, k**). **l-o,** Calculated TSS dispersions at the side-surface (100) for the WTI phase (**l, m**) and the STI phase (**n, o**) along distinct high-symmetry lines (red and yellow lines in **e**).

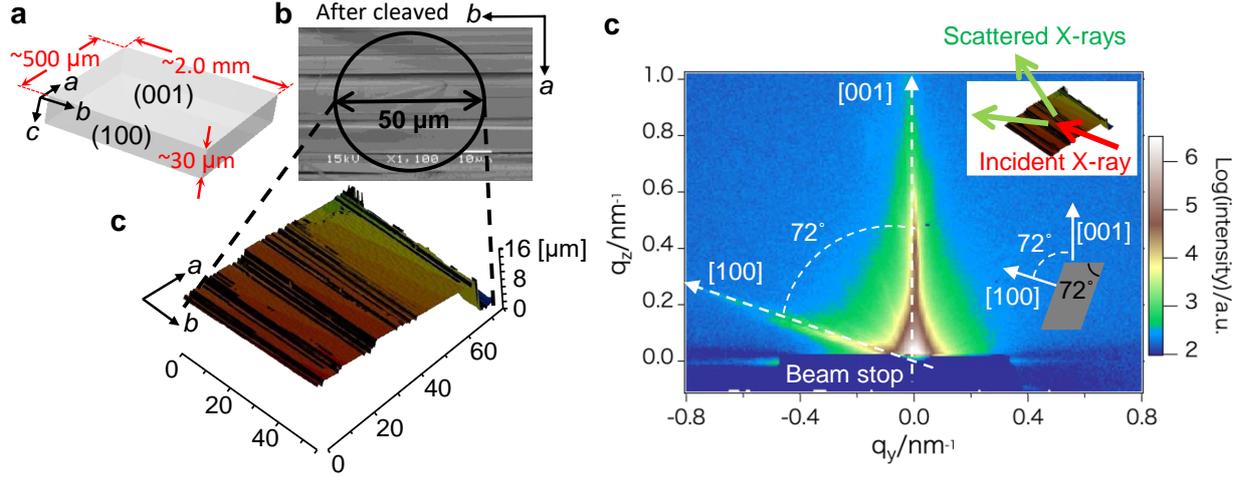

FIG. 2: **Two cleavage surfaces of $\beta$-Bi$_4$I$_4$ a,** Typical dimensions of $\beta$-Bi$_4$I$_4$ samples. **b,** SEM and **c,** laser-microscope image of a cleaved surface, respectively. Black circle indicates the typical spot size of laser light used in our ARPES experiment (see Fig. 3). **d,** Result of grazing-incidence small angle X-ray scattering (GISAXS) measurement of a cleaved $\beta$-Bi$_4$I$_4$ after ARPES measurement. The chain direction of the sample ($b$ axis) was aligned parallel with the incident X-ray.

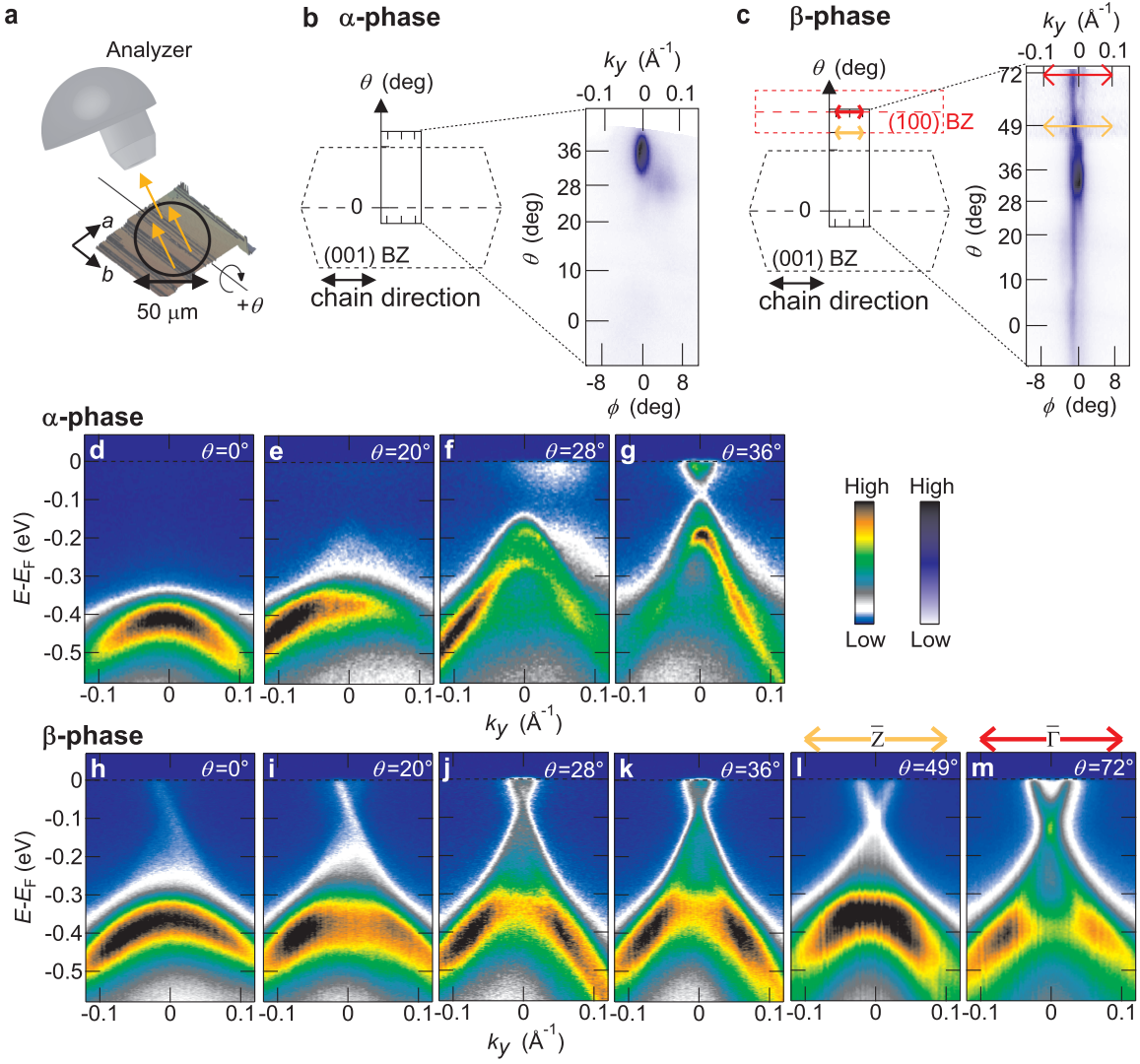

FIG. 3: **Experimental band structures of $\alpha$- and $\beta$-Bi$_4$I$_4$ by laser-ARPES. a,** Experimental geometry of laser-based ARPES. We collect the photoelectrons at different emission angle $\theta$ with respect to the normal of the top-surface (001). **b, c,** Photoelectrons intensity distributions at $E_F$ for the two materials. The intensities are integrated within 40 meV. The black and red dashed lines display the surface Brillouin zones of (001) and (100) surfaces, respectively. **d-g** and **h-m**, APRES band maps taken at different $\theta$-s in the $\alpha$-phase and the $\beta$-phase, respectively. In **f** and **j**, we find that the top of the valence bands split in energy in the $\alpha$-phase differed from those in the $\beta$-phase. This can be assigned to the bilayer splitting according to our DFT calculations (Supplementary information). In **l** and **m**, ARPES band mapping along the high-symmetry lines cut at the $\bar{Z}$ (red arrows in **c**) and $\bar{\Gamma}$ (yellow arrow in **c**) of the (100) surface BZ.

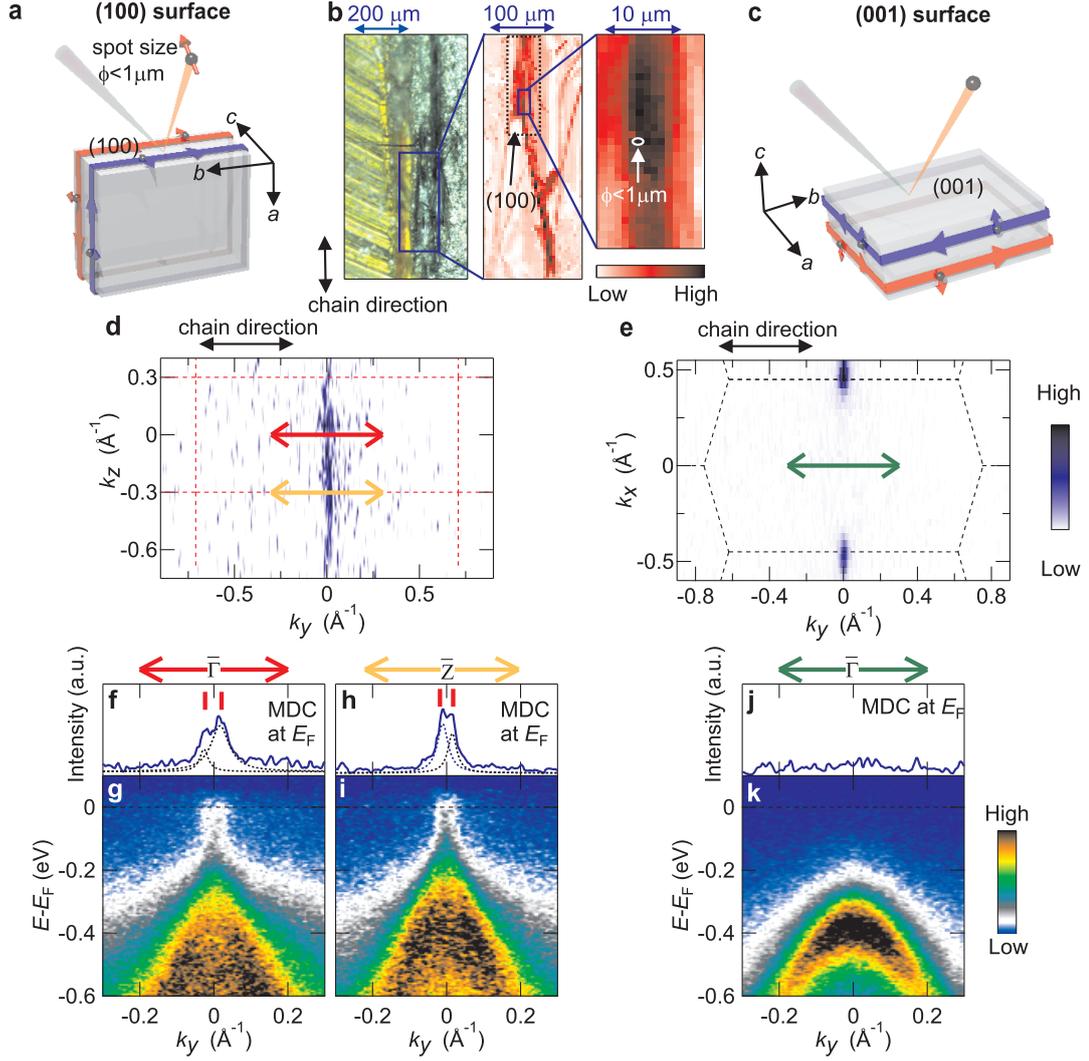

FIG. 4: **Surface-selective nano-ARPES of $\beta$-Bi$_4$I$_4$ with $h\nu$=85 eV. a,** Schematic geometry for surface-selective non-ARPES at the side-surface (100) of $\beta$-Bi$_4$I$_4$. **b,** Real space images of the measured sample taken by a optical microscope (left), the photoemission intensity maps (middle) and the magnified image (right). The white circle indicates the position where we perform the surface-selective experiment. **c,** Schematic of different surface-selective nano-ARPES at the top-surface (001). **d, e** Photoelectron intensity distributions at $E_F$. The intensities are integrated within 100 meV, and collected without the sample rotation by using a capability of the electron deflector of the photoelecton analyzer (Methods). The dashed lines denote the surface Brillouin zone of (red) the (100) surface (**d**) and (black) the (001) surface (**e**). **f-i,** ARPES band maps and their momentum distribution curves (MDCs) at the $E_F$ cut the $\bar{\Gamma}$ and $\bar{Z}$ points of the (100) surface. The black dotted lines are the Lorentzian curves fitting the two peak structures in the MDCs. Their peak positions are indicated by the red lines in **f** and **h**. **j, k,** ARPES images around the $\bar{\Gamma}$ point of the (001) surface (**m**) and the MDC at the $E_F$ (**j**).

18